\documentclass[5p, twocolumn]{elsarticle}
\usepackage{epsfig,amssymb,amsfonts,bm,booktabs}
\biboptions{sort&compress}

\usepackage[utf8]{inputenc}
\usepackage[english]{babel}
\usepackage[dvipsnames]{xcolor}
\usepackage[normalem]{ulem}
\usepackage[
colorlinks=true,
linkcolor=blue,
breaklinks=true,
urlcolor=magenta,
citecolor=blue]{hyperref}

\usepackage{bm,color,xcolor,amsfonts,amsmath,amssymb,epsfig,esint,orcidlink}
\usepackage{cancel}

\newcommand{\0}{^{\rm{ph}}}

\newcommand{\be}{\begin{equation}}
\newcommand{\ee}{\end{equation}}
\newcommand{\bea}{\begin{eqnarray}}
\newcommand{\eea}{\end{eqnarray}}
\newcommand{\beas}{\begin{eqnarray*}}
\newcommand{\eeas}{\end{eqnarray*}}

\newcommand{\vep}{{\bm p}} 

\newcommand{\veq}{{\bm q}}

\newcommand{\mpi}{m_{\pi}}

\newcommand{\lhc}{{\rm lhc}}

\newcommand{\B}{\color{blue}}

\begin{document}

\begin{frontmatter}

\title{Internal structure of the $T_{cc}(3875)^+$ from its light-quark mass dependence}

\author[1]{Michael Abolnikov\orcidlink{0009-0006-9874-0876}}
\author[1]{Vadim Baru\orcidlink{0000-0001-6472-1008}}
\author[1]{Evgeny Epelbaum\orcidlink{0000-0002-7613-0210}}
\author[1]{Arseniy A. Filin\orcidlink{0000-0002-7603-451X}}
\author[2]{Christoph Hanhart\orcidlink{0000-0002-7603-451X}}
\author[1]{Lu Meng\orcidlink{0000-0001-9791-7138}}

\address[1]{Institut f\"ur Theoretische Physik II, Ruhr-Universit\"at Bochum, D-44780 Bochum, Germany}
\address[2]{Institute for Advanced Simulation (IAS-4), Forschungszentrum J\"ulich, D-52425 J\"ulich, Germany}

\begin{abstract}
  We employ a chiral effective field theory-based approach to connect
  $DD^*$  scattering observables at the physical and variable pion
  masses accessible in lattice  QCD simulations.  
  We incorporate all relevant scales associated with
  three-body $DD\pi$ dynamics and the left-hand cut
induced by the one-pion exchange for pion masses
higher than the physical one, as required by analyticity and
unitarity. By adjusting the contact interactions to match experimental
data at the physical pion mass and lattice finite-volume energy levels
at $m_{\pi} = 280$ MeV, we predict the trajectory of the $T_{cc}$ pole
as a function of the pion mass, finding it consistent with the
hadronic-molecule scenario. In particular, we find that the explicit
treatment of the one-pion exchange has a pronounced effect on the pole
trajectory for $\mpi \gtrsim 230$ MeV by pushing it into the complex energy plane.

\end{abstract}

\end{frontmatter}


\section{Introduction}

The spectroscopy of mesons and baryons containing hidden and open heavy flavor quarks has made tremendous progress in recent years.
Experimental observations have provided evidence for numerous
multiquark exotic states, including tetraquarks for mesons and
pentaquarks for baryons, with many of them being found in close
proximity to certain hadron-hadron thresholds, as summarized in recent
reviews~\cite{Esposito:2016noz,Lebed:2016hpi,Guo:2017jvc,Yamaguchi:2019vea,Brambilla:2019esw,Guo:2019twa,Chen:2022asf,Meng:2022ozq}. However,
the structure of these exotic hadrons, determined by the internal
clustering of quarks, is largely unknown and remains the subject of extensive research.   
Prominent theoretical scenarios  for these structures include hadronic
molecules,  compact multiquark states and atom-like hadronic
configurations called hadroquarkonia,
where a compact heavy quarkonium serves as a core.
The size of a molecule made out of two
hadrons is  controlled by the inverse binding momentum $\gamma
=\sqrt{2\mu E_B}$, where $E_B$ denotes the binding energy and $\mu$
the reduced mass of the
hadrons. Thus, for very small binding
energies, hadronic molecules acquire a very large size: The state of interest for this paper has $E_B\approx
300$~keV, yielding a spatial extension of the order of $1/\gamma\approx
8$ fm. Conversely, compact states are characterized by a size of the
order of $1/\Lambda_{\rm QCD}\simeq 1$~fm, where $\Lambda_{\rm QCD}$
represents the typical scale of the strong interactions.

In 2021, the LHCb experiment discovered 
the first exotic doubly-charmed narrow resonance, denoted as $T_{cc}(3875)^+$, whose minimal quark composition is $cc\bar u \bar d$~\cite{LHCb:2021vvq,LHCb:2021auc}. 
This discovery revealed a state with a mass just a few hundred keV
below the $D^{*+}D^0$ threshold, with a width primarily dictated by its strong decay mode to $DD\pi$. 
 Since approximately 90\% of the $D^0D^0\pi^0$ events contain a genuine $D^{*}$ meson~\cite{LHCb:2021auc}, 
 it is natural to expect (see the discussions in Refs.~\cite{Feijoo:2021ppq,Yan:2021wdl,Fleming:2021wmk}) that the width of the $T_{cc}(3875)^+$ should be smaller than that of the $D^{*}$, which is only $(83.4 \pm 1.8)$ keV \cite{ParticleDataGroup:2024cfk}. 
The properties of this state have been investigated 
 employing low-energy effective field theories (EFT)
 \cite{Albaladejo:2021vln,Meng:2021jnw,Du:2021zzh,Braaten:2022elw,Wang:2022jop,Dai:2023mxm}
 and phenomenological models, see, e.g.,  \cite{Chen:2022asf} and
 references therein.  In particular,  the pole position of the
 $T_{cc}(3875)^+$ and the $DD^*$
 scattering parameters were extracted in Ref.~\cite{Du:2021zzh} using 
 the leading order chiral EFT approximation by performing
 a coupled-channel  analysis of the experimental line shape  in the   $D^0D^0\pi^+$ final state.   In that
work, a special attention was paid  to    the inclusion of three-body
cuts,  which were found to be crucial for the accurate determination of the $T_{cc}$ pole position in the complex energy plane. 
We emphasize that for the $T_{cc}$, including the one-pion exchange (OPE) to all orders with a proper treatment of its cuts is needed for theoretical consistency. To illustrate this, we note that the $D^*$ self-energy induces an imaginary part to the $D^*D$ propagator, which corresponds to the $DD\pi$ intermediate state going on shell. This on-shell intermediate state contains a pair of identical mesons ($DD$) in isospin 1, which has to  be in an even partial wave to comply with Bose symmetry. Proper symmetrization of the $DD$ state, that ensures this, generates a one-pion exchange. Moreover, since the $D^*$ self-energy is resummed in the $D^*D$ propagator, consistency with Bose symmetry requires resumming the OPE as well. For a similar discussion on the Pauli principle in the context of two identical nucleons in the three-body intermediate state, see Refs.~\cite{Baru:2004kw,Lensky:2005hb,Baru:2009tx}.
 
More recently, $DD^*$ scattering has also been investigated in lattice QCD employing the L\"uscher method~\cite{Padmanath:2022cvl,Chen:2022vpo,Whyte:2024ihh,Collins:2024sfi}
and the HAL QCD approach~\cite{Lyu:2023xro}.  In particular, the analysis performed in Ref.~\cite{Padmanath:2022cvl}
by utilizing the effective range approximation (ERE) suggests that the
$T_{cc}$ state is consistent with a virtual state at the pion mass
$m_{\pi}=280$ MeV.  
However, two significant concerns have been raised in the literature
regarding this conclusion.
First, it was emphasized in  Ref.~\cite{Du:2023hlu} that the ERE is
only valid for parameterizing the near-threshold energy behavior of
the inverse scattering amplitude if there are no nearby left-hand cuts (lhc) generated by long-range interactions. 
The presence of the OPE in the $DD^*$ scattering
potential introduces a left-hand branch point at the center-of-mass momentum
$|p_{\rm lhc}^{1\pi}|=126$~MeV for the given pion mass, thereby
significantly restricting the applicability of the effective range
expansion.
The second issue concerns the L\"uscher method that is widely used for
extracting infinite-volume scattering amplitudes from finite-volume
energy levels calculated on the
lattice~\cite{Luscher:1986pf,Luscher:1990ux,Kim:2005gf,Briceno:2014oea},
see also
\cite{Gockeler:2012yj,Leskovec:2012gb} for generalizations to moving two-body systems and 
\cite{Aoki:2020bew,Mai:2021lwb,Bicudo:2022cqi,Bulava:2022ovd,Prelovsek:2023sta} for recent reviews. 
Recent studies have argued that this method faces difficulties in situations involving nearby left-hand cuts~\cite{Raposo:2023oru,Dawid:2023jrj,Green:2021qol,Meng:2023bmz}.
Consequently, several extensions of this method or alternative approaches have been proposed to address this issue~\cite{Raposo:2023oru,Meng:2023bmz,Bubna:2024izx,Hansen:2024ffk}. In particular, in Ref.~\cite{Meng:2023bmz}, a solution of the lhc problem was proposed based on the chiral EFT approach. Due to the explicit account for the longest-range interaction from the OPE in this approach, 
finite-volume energy levels can be directly calculated as solutions of the eigenvalue problem both below and above the left-hand cut.  	
The distinctive feature of Ref.~\cite{Meng:2023bmz} is that it not only provided an alternative to the L\"uscher method, which is valid
in the presence of the lhc, but also allowed, for the first time, to
take into account the left-hand cut effects in the
analysis of the actual lattice data.
Using the 
lattice energy levels from  Ref.~\cite{Padmanath:2022cvl} and
taking into account the lhc effects, Ref.~\cite{Meng:2023bmz} found the  $T_{cc}$ to have a
$85\%$ probability of being a  resonance state (based on the
scattering amplitude calculated at the $1\sigma$ confidence level), 
while the remaining $15\%$ probability corresponds to a scenario with
two virtual poles~\cite{Meng:2023bmz}.  The single virtual pole
extracted in Ref.~\cite{Padmanath:2022cvl} is incompatible with the
presence  of the left-hand cut, as was already shown in Ref.~\cite{Du:2023hlu}.
 
In this work,  we employ
the same chiral EFT-based approach to scrutinize the analytic structure of the scattering amplitude as a function of the pion mass. 
Given the intricate interplay between the right-hand (three-body) and
left-hand cuts, which depend sensitively upon the pion mass, we incorporate all energy and/or momentum scales relevant
for this complicated dynamics to predict the pole trajectory of the  $T_{cc}$ state for 
pion masses between  the physical value $m_{\pi}^{\rm ph}$ and  $3 m_{\pi}^{\rm ph}$.  This information
can serve as a benchmark for  future lattice QCD calculations and is
important for obtaining additional insights into the structure of
$T_{cc}$ state~\cite{Matuschek:2020gqe}.  
Our analysis also yields predictions for the $DD^*$ phase shifts at any value of  $m_{\pi}$
within the considered range, in spite of the fact that 
the ERE has a very limited range of validity. 
The  lhc also necessitates an improvement of the Weinberg approach to
properly describe the compositeness of a hadronic state. Besides the fact that
the zeros in the $T$-matrix emerging from the interplay of the
repulsive OPE potential and attractive short-range physics may
invalidate the original Weinberg formalism, as discussed in
Refs.~\cite{Weinberg:1965zz,Baru:2010ww,Kang:2016jxw}, already the
small scale introduced into the system by the  nearby lhc calls for a 
refined formulation of the compositeness criterion.

\section{Framework}
\label{Sec:framework}

In this work, we use  both experimental data~\cite{LHCb:2021vvq,LHCb:2021auc} and lattice energy levels~\cite{Padmanath:2022cvl} to determine the a priori unknown low-energy constants in chiral EFT. With these fixed, we 
 predict the $DD^*$ scattering amplitude at various pion masses. Before discussing the explicit procedure, several remarks are in order:
 \begin{itemize}
       \item  We focus on observables near  the $DD^*$ threshold, so we do not consider potential coupled-channel effects involving $D^*D^*$.  
        In a very recent lattice 
investigation of coupled-channel  $DD^*-D^*D^*$ scattering at $\mpi =
391$ MeV,  a sizeable coupled-channel effect was
reported~\cite{Whyte:2024ihh}  within the L\"uscher formalism using,
however, amplitude 
parameterizations that ignore the left-hand cuts.  We briefly comment
on these results below.  However,  since no information about the
coupled-channel dynamics is available from
Ref.~\cite{Padmanath:2022cvl}, which is used  as input 
for our study,  these effects are ignored.        
       \item  Our calculations are performed in the isospin limit using the averaged masses 
       $M_{D^{(*)}} = (M_{D^{(*)0}}+M_{D^{(*)c}})/2$ for the $D^{(*)}$-mesons.  
       Additionally, since electromagnetic effects are not yet resolved on the lattice, we ignore the radiative decay width of the $D^*$,  considering only its strong pionic decay.      
       \item 
The main goal of this study is to analyze the pion mass dependence of $DD^* $ scattering observables in the continuum limit. The lattice energy levels used here as input  
are obtained at a single lattice spacing of $a \approx 0.086 \, \text{fm} $. 
Previous investigations, such as Ref.~\cite{Junnarkar:2018twb}, found
the dependence of finite volume energy levels on the lattice spacing
in the doubly heavy sector to be moderate for the $ud\bar{c}\bar{c}$ 
case. Based on these findings, we neglect the lattice spacing dependence in our analysis.
\end{itemize}

The effective potential $V$ for $DD^*$ scattering is constructed in chiral EFT up to ${\cal O}(Q^2)$, where 
$Q=p/\Lambda_b$  with $p \sim m_{\pi}$  being a characteristic soft momentum scale and $\Lambda_b$ referring to the breakdown scale of the chiral expansion,   
and is given by
\begin{equation}
V=V_{\text{OPE}}^{(0)}+V_{\text{cont}}^{(0)}+V_{\text{cont}}^{(2)}+...\ .\label{eq:veft}
\end{equation}
Here, we assume that the two-pion exchange contributions 
are largely saturated by the contact terms, see also Refs.~\cite{Baru:2013rta,Baru:2015tfa} for related studies in the context of the $X(3872)$.

In analogy to the $NN$ system~\cite{Hammer:2019poc,Epelbaum:2019kcf}, it
was shown in Ref.~\cite{Baru:2015nea} that the OPE potential in
heavy-meson systems is well defined, in the EFT sense, only in combination
with contact operators. These contact terms 
account for our ignorance of short-range dynamics
and have the form of a polynomial function in the pion mass
and momenta.
The   isoscalar contact potentials contributing to the relevant $^3
S_1$\footnote{Here and in what follows, we use the spectroscopic
  notation $^{2S+1}L_J$ to indicate a $DD^*$ partial wave  with total spin $S$, angular momentum $L$ and
total spin equal to $J$.}
partial wave near the $DD^*$ threshold 
can be parametrized as
\begin{equation}\label{full_NLO_contact_ansatz}
    \begin{split}
         & V_{\text{cont}}(p,p^\prime) = \left[c_0(\xi) + c_2(\xi)(p^2+p^{\prime 2}) \right]  ({\bm\epsilon \cdot \bm\epsilon'^*}  ) \ , \\
           & c_0(\xi) = C_0+D_2(\xi^2-1)+\mathcal{O}(\xi^4, p^4)\ , \\
        & c_2(\xi) = C_2 +\mathcal{O}(\xi^2)\ , 
    \end{split}
\end{equation}
where $\xi=m_{\pi}/m_{\pi}^{\rm ph}$ while $\bm p$ $(\bm p')$ and $\bm\epsilon$ $(\bm\epsilon')$ 
denote the center-of-mass momentum  and   polarization 
of the initial (final) $D^*$ meson, respectively.
The values of the low-energy constants (LECs) in
Eq.~\eqref{full_NLO_contact_ansatz} are
determined from 
empirical data at the physical pion mass ($\xi=1$) as well as from the
lattice energy levels at certain $\xi$-values away from the physical point, as discussed in Sec.~\ref{Sec:pole}.  
In this work, for this purpose, we use the lattice QCD data at
$m_{\pi} = 280$ MeV, which corresponds to $\xi \approx 2$.  The effect
of  higher-order contact terms at $\mathcal{O}(Q^4)$ is estimated in
Sec.~\ref{sec:theor_uncert}. 

In the framework of time-ordered-perturbation theory (TOPT),  the isoscalar  OPE potential is given by 
\bea
\hspace{-0.5cm}V_\text{OPE}(E,\vep,\vep')=-\frac{g^2}{8f_\pi^2} \frac{({\bm q\cdot \bm\epsilon}  )  ({\bm q \cdot \bm\epsilon'^*  })}{2\omega_\pi(\veq^2)}\, D^\pi(E,\vep,\vep'), 
\label{OPEfull}
\eea
where $f_\pi$ is the pion decay constant  and $g$  is the coupling constant of pions with heavy mesons. Furthermore,
\be\label{Eq:Dprop}
D^\pi(E,\vep,\vep') ={D_1(E,\vep,\vep') +D_2(E,\vep,\vep') },
\ee
and the two TOPT propagators read
\bea
\hspace{-0.5cm}D_1(E,\vep,\vep') &\!\!=\!\!&\left(2M_D+\frac{p^2+p'^2}{2M_D}+\omega_\pi(\veq^2)-E-i\epsilon\right)^{-1}, \nonumber\\
\hspace{-0.5cm}D_2(E,\vep,\vep') &\!\!=\!\!&\left(2M_{D^*}+\frac{p^2+p'^2}{2M_{D^*}}+\omega_\pi(\veq^2)-E-i\epsilon\right)^{-1}.\nonumber
\eea
Here, $E$ is total energy of the system and  $\veq {=}\vep{+}\vep'$.
  We treat    the pion  relativistically, such that $\omega_\pi(\veq^2){=}\sqrt{m_\pi^2{+}\veq^2}$,  while the $D^{(*)}$ mesons are nonrelativistic. 
The resulting potential can be therefore written as 
\bea
V(E,\vep,\vep')=V_{\text{cont}}(p,p^\prime)+V_\text{OPE}(E,\vep,\vep'). 
\eea
 The partial-wave-projected  potentials  $V_{\alpha\beta}(E,p,p')$ are obtained along the lines of Refs.~\cite{Baru:2016iwj,Baru:2019xnh} as follows, 
\bea
V_{\alpha\beta}(E,p,p')&=&\frac1{2J+1}\int\frac{d \Omega_p}{4\pi}\frac{d \Omega_{p'}}{4\pi}\\ 
\nonumber \times\ {\rm Tr}\Bigl[P^\dagger(JLS;\bm n) \hspace{-0.6cm}&&\hspace{-0.3cm} V(E, \bm p,\bm p')  P(JL'S';\bm n')\Bigr],
\label{VPWA}
\eea
where the Greek indices run from 1 to 2 accounting for the $^3S_1$ and $^3D_1$ partial waves, respectively, with
$L(L')=S$ or $D$; $\bm n=\bm p/p$ ($\bm n'=\bm p'/p'$), and a complete set of relevant properly normalized 
projection operators $P(JLS;\bm n)$ is given in the Appendix of Ref.~\cite{Baru:2019xnh}.

The  scattering amplitude is calculated as a solution of the Lippmann-Schwinger equation
\bea
\label{eq:lse}
T_{\alpha\beta}(E,p,p') &=& V_{\alpha\beta}(E,p,p')\\
& &\hspace{-2.cm} +  
 \int_0^\Lambda   \frac{\text{d}{q}\, q^2}{2\pi^2}  
V_{\alpha\gamma} (E,{\B p},q)G(E,q)T_{\gamma\beta}(E,q,{\B p'}).\nonumber
\eea
In order to render the integral in \eqref{eq:lse} well defined, we use a sharp cutoff regularization. Specifically, the main results were obtained using  
$\Lambda=700$~MeV, but for the purpose of testing the cutoff dependence we  also use $\Lambda=500$~MeV.
The $DD^*$ propagator is expressed as 
\bea
G(E,q) = \left[M_{D^*}+M_D+\frac{q^2}{2\mu}-E-\frac{i}{2}\Gamma(E,q)\right]^{-1},~~
\eea
where $\mu = M_D M_{D^*}/(M_D+M_{D^*})$ is the reduced mass,
$$
\Gamma(E,q) = \frac{g^2M_D}{8\pi f_\pi^2 M_{D^*}}\Big[\Sigma(s) - \Sigma_0(s)\theta(M_D+m_\pi-M_{D^*}) \Big]
$$
 is the dynamical width of the $D^*$ with 
\bea
\Sigma(s) = \left[ \frac{\sqrt{\lambda(s,M_D^2,m_\pi^2)}}{2\sqrt{s} }\right]^{3} ,
\eea
and $\lambda(a,b,c)=a^2+b^2+c^2-2ab-2bc-2ca$ is the K\"all\'en triangle function, and $s=[E-M_D-q^2/(2\mu)]^2$. Here
\bea
\Sigma_0(s)&=& \Sigma(M_{D^*}^2)\nonumber\\
\hspace{-0.6cm}&&\hspace{-0.5cm}+ 2M_{D^*}\left(E-M_{D^*}-M_D-\frac{q^2}{2\mu}\right)\Sigma'(M_{D^*}^2),\nonumber
\eea
where the first and second terms renormalize the $D^*$ mass and wave function, respectively, if $M_{D^*}<M_D+m_\pi$.
The dependence of the $D$- and $D^*$-meson masses on $\mpi$ was explored in Ref.~\cite{Cleven:2010aw} using unitarized SU(3) chiral perturbation theory. 
The tree-level expressions for $M_{D}$ and $M_{D^*}$, derived from the expansion around the physical masses 
with the charm quark mass kept at its physical value, are as follows, 
\bea
M_{D}(\xi)&=&M_{D}\0\left[1+h_1\left(\frac{m_\pi\0}{M_{D}\0}\right)^2(\xi^2-1)\right],\quad \\
M_{D^*}(\xi)&=&M_{D^*}\0\left[1+h_1\left(\frac{m_\pi\0}{M_{D^*}\0}\right)^2(\xi^2-1)\right],\quad
\label{mDs}
\eea
where $h_1\approx 0.42$.  
The pion mass dependence of the pion decay constant $f_\pi$ and the
$\pi DD^*$ coupling constant $g$ is considered along the lines of
Refs.~\cite{Du:2023hlu,Becirevic:2012pf} --- see Sec.~3 of the Supplemental Material in Ref.~\cite{Du:2023hlu}.  Note also that  the extraction of the  coupling constant $g$ 
was recently updated in Ref.~\cite{Du:2023hlu}  by making 
 two-dimensional fits of lattice data~\cite{Becirevic:2012pf} with  
simultaneously varied  $m_{\pi}$ and the lattice spacing.  We have verified that  employing the updated coupling has a small impact on the results lying well within the theoretical uncertainty estimate 
discussed in  Sec.~\ref{sec:theor_uncert}.
We therefore do not dwell on this any further.

\section{Pole position as a function of the pion mass}  
\label{Sec:pole} 

\subsection{{\rm LO} results}\label{sec:pole_LO}

\begin{figure*}[t]
    \centering    
      \includegraphics[width=0.4\linewidth]{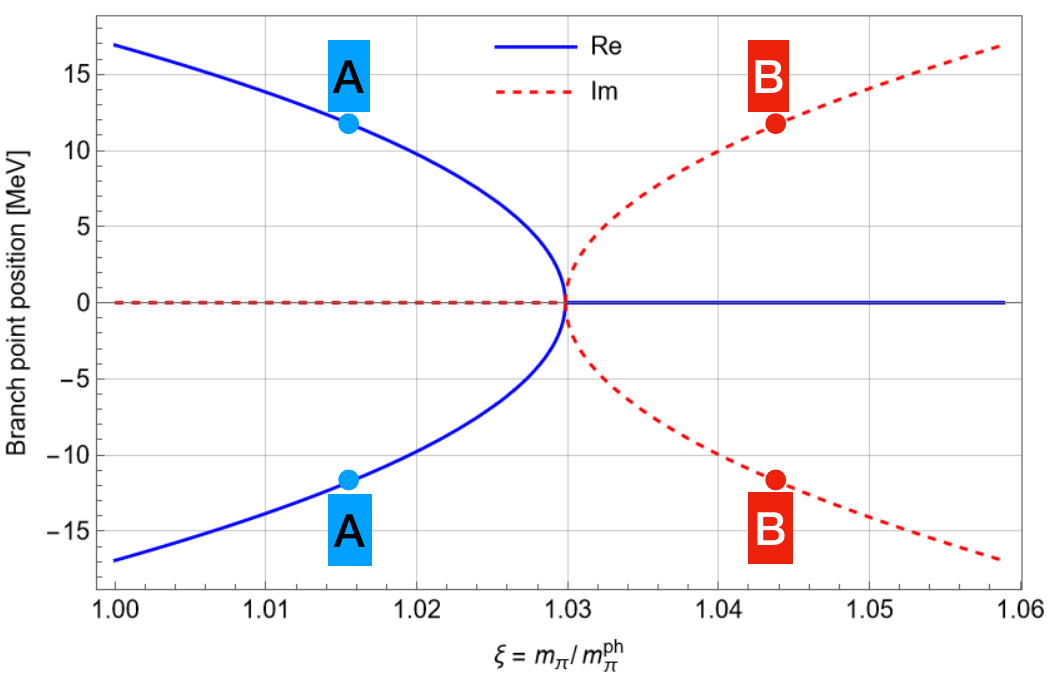} \hspace{0.03\textwidth}  
\raisebox{0.75cm}{\includegraphics[width=0.54\linewidth]{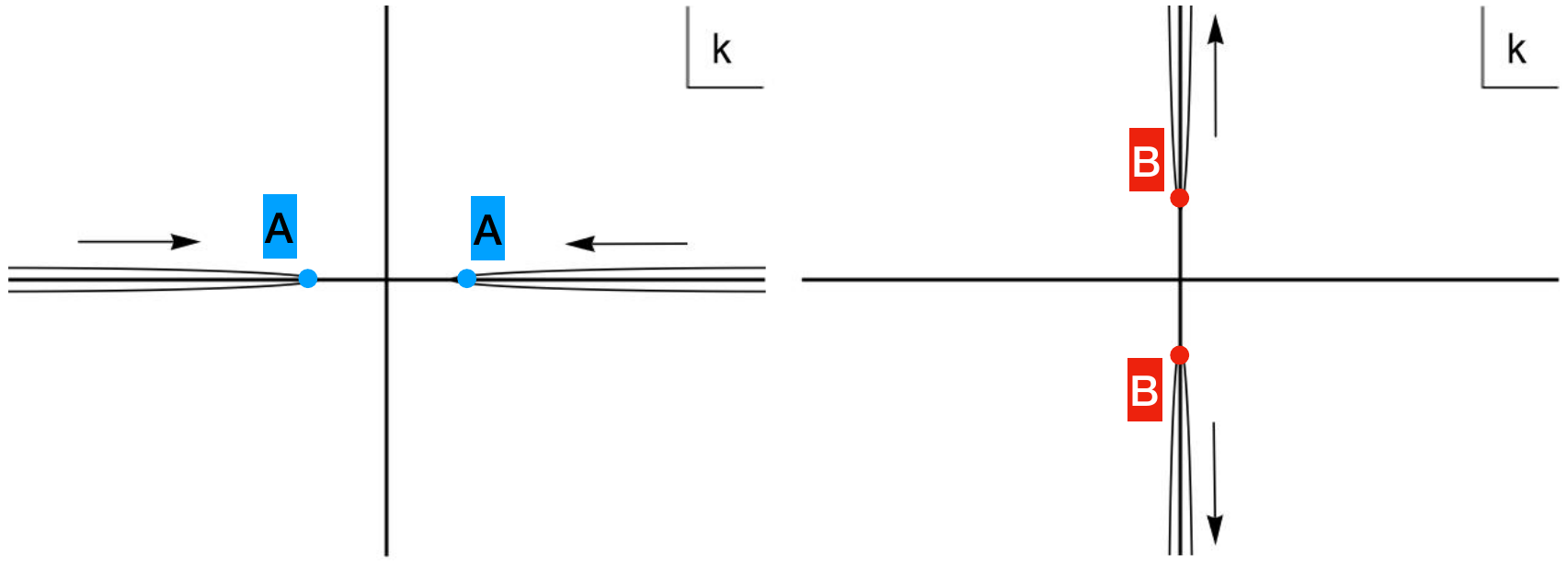}}  
           \caption{Branch cuts in the on-shell partial-wave projected OPE potential  versus  the pion mass.  
          Left panel:  Re (blue solid lines) and Im (red dashed lines) parts of the branch point momentum.  Real momenta correspond to a three-body cut,  imaginary ones to a lhc.      
        At $\xi = \xi_0 \approx 1.03$, the right-hand cut   changes into a left-hand cut. 
        Middle and right panel:  Schematic behavior of  the branch cuts in the complex $k$ plane when the pion mass is increased.  The direction of change is indicated by arrows.
        The blue points A indicate the branch points of the right-hand cut at some value of $\xi \in [1, 1.03]$.  The red points B  indicate the branch points of the left-hand cut at some value of $\xi >  1.03$.
}
   \label{moving_branch_cuts}
\end{figure*}

\begin{figure*}[t]
    \centering    
\vspace*{0.7cm}
\raisebox{0.75cm}
{\includegraphics[width=1\linewidth]{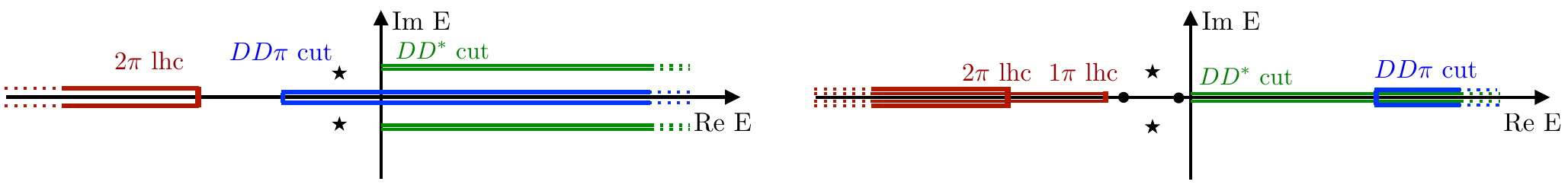}}  
\vspace*{-1.cm}
           \caption{Sketch of the locations of various branch cuts and
poles in the complex energy plane for 
 the physical pion mass (left panel) and $m_\pi$ = 280 MeV (right panel). The left-hand cut, the two-body $D^*D$ cut,
and the three-body $DD\pi$ cut are shown in red, green, and blue,
respectively. The black symbols show typical locations for
the $T_{cc}^+$ poles which show up either as a pair of virtual states
(dots) or a resonance (stars) in case for the larger pion mass or as quasi-bound state at the physical pion mass. For the two scenarios the poles are located on the second and first Riemann sheet
with respect to the $DD^*$ cut, respectively.
}
   \label{ampl_cuts_E}
\end{figure*}

Before discussing the next-to-leading order (NLO) results in the following section, it is worth noting that all pole trajectories for this problem exhibit several common features, which we 
can illustrate using the leading-order (LO) results. In chiral  EFT at LO, there is only one contact term,  $C_0$
 (see Eq.~\eqref{full_NLO_contact_ansatz})  which is adjusted  to reproduce the real part of the 
 $T_{cc}$  pole position, ${\rm Re} E_{\rm pole}=-356$ keV, extracted
 in Ref.~\cite{Du:2021zzh} (see "pionful fit III" in Table II) from a
 chiral EFT-based analysis of the experimental data.  The imaginary
 part of  the $T_{cc}$ pole at the physical pion mass is  governed by
 its three-body decay to $DD\pi$ and, therefore, comes out as a
 prediction since the $D^*D\pi$ coupling is known. 
 Then, the  $T_{cc}$ pole trajectory  for pion masses other than the physical one is predicted based on the interplay of several scales, primarily associated with the OPE, as discussed below.

\indent The analytic structure of the scattering amplitude in the complex momentum plane ($k$-plane)  is continuously changing with varying pion mass.  At the nominal $D^*D$ threshold,
{the OPE has both a real and an imaginary part as long as the
  three-body $DD\pi$  threshold is below the $DD^*$ threshold. This is  fulfilled, in particular, 
for the physical pion mass ($\xi=1$). }
Defining the on-shell momentum relative to the two-body threshold by  $k$, with $E=M_D +M_{D^*} + k^2/(2\mu)$, and introducing $\Delta M=M_{D^*}-M_D $,  the three-body branch point can be found 
by requiring $E_{\rm rhc_3}{=M_D +M_{D^*} + k^2_{\rm rhc_3}/(2\mu)} = 2 M_D + m_{\pi}$.  This leads to
\be\label{Eq:branch_3body}
k^2_{\rm rhc_3}= 2\mu (m_{\pi}-\Delta M).
\ee
This relation can also be derived  by setting $p=p' =0$ in the propagator $D_1$ in  Eq.~\eqref{Eq:Dprop},  resulting in  $k^2_{\rm rhc_3} < 0$ and $|k_{\rm rhc_3}|\approx 88$ MeV at the physical pion mass. 
Even when  $p$ and $p'$ are different from  zero, the cut  can still occur at the given  pion mass, if the denominator in $D_1$ vanishes.  However,  in these cases, it will emerge at $k^2> k^2_{\rm rhc_3}$.

 When both $p$ and $p'$ are on shell ($p=p' = k$) and $m_{\pi}{>}\Delta M$,  
the OPE and, consequently, the on-shell $DD^*$ partial wave amplitudes, exhibit the left-hand cut
at imaginary values of the momenta.  The lhc branch point closest to the threshold is given by   \cite{Du:2023hlu}
\be
(k_\lhc^{1\pi})^2\approx\frac14[(\Delta M)^2-m_\pi^2].
\label{plhc}
\ee
This can be derived from $D_1$ using that $\mu\approx m_D/2$, leading to the conclusion that  $\omega (4k^2)=\Delta M$ for  forward $DD^*$ scattering ($\cos{(\vep\cdot\vep')}=1$).

\begin{figure*}[t]
    \centering    
    \begin{minipage}{0.52\textwidth}
      \includegraphics[width=1\linewidth]{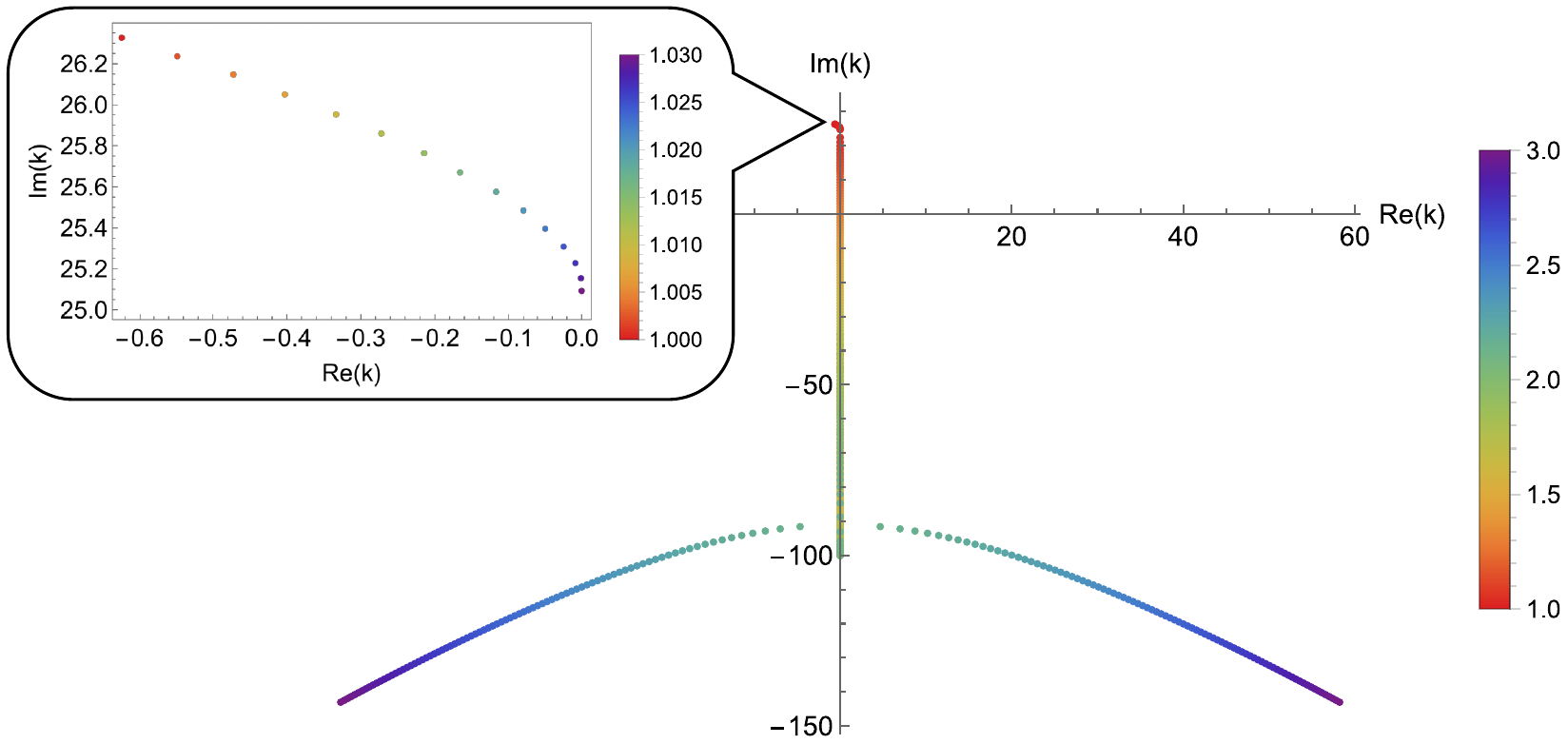} \hspace{0.05\textwidth} 
      \end{minipage}
       \hspace{0.03\textwidth}
 \begin{minipage}{0.43\textwidth}
       \includegraphics[width=\linewidth]{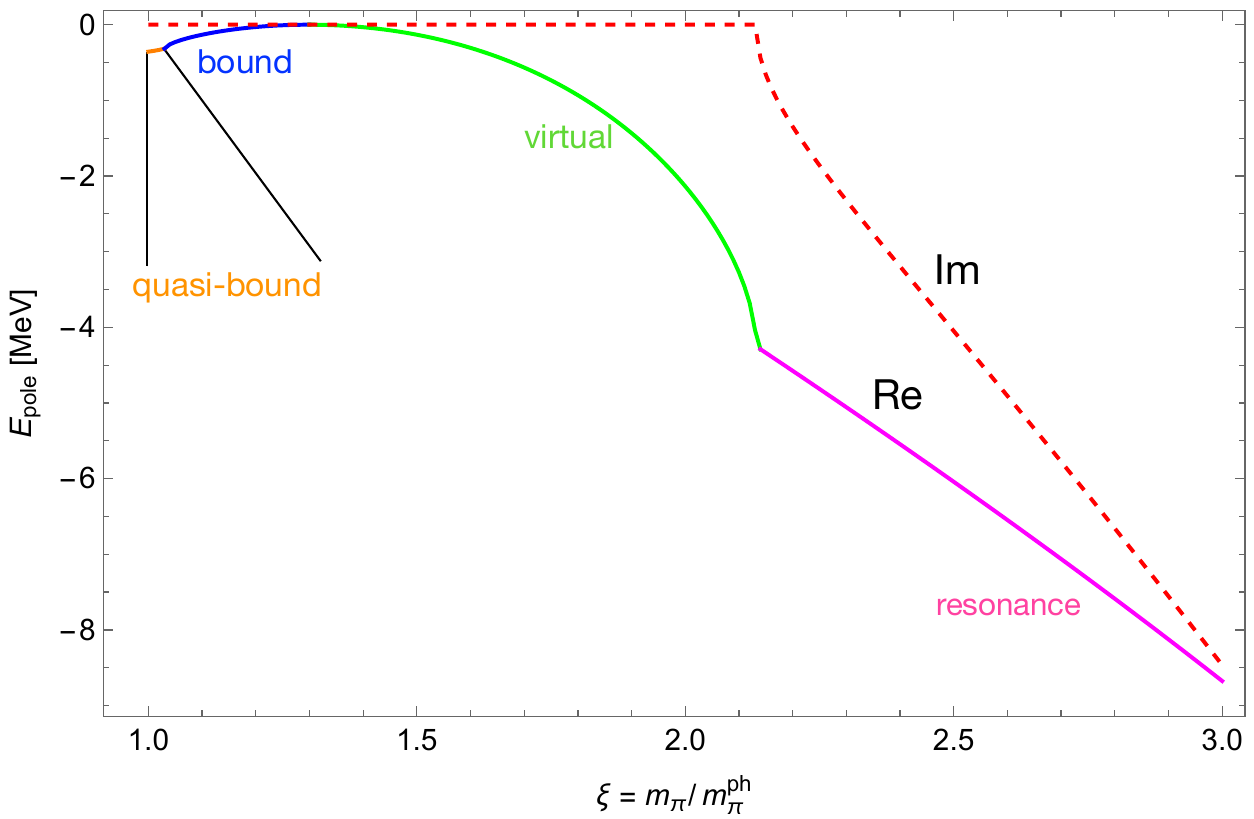}  
       \end{minipage}
        \caption{
         Left panel: Pion mass dependence of the $T_{cc}$ pole in the complex $k$-plane predicted at LO in chiral EFT. The value of $\xi$ is indicated by color.   The second, more distant virtual state is also shown, along with its collision point with the $T_{cc}$ pole, after which the  resonance poles emerge.  
            Right panel:   LO  trajectory  of the $T_{cc}$ pole  in the complex energy plane as a function of $\xi$, corresponding to the left panel.  Re $E_{\rm pole}$ and  Im $E_{\rm pole}$ are shown by solid and dashed lines, respectively.       
          The second, more distant  state is not shown;   only the energy of the pole with $\text{Re}(k)>0$ is shown
           after the resonance appears.}
   \label{LO_pole_trajectory}
\end{figure*}
The partial-wave projected OPE potential with both initial and final $D^*D$ pair on-shell
 can exhibit either a lhc or a three-body cut, depending on the pion mass, as illustrated in Fig.~\ref{moving_branch_cuts}. 
When the pion mass increases from its physical value, the three-body
phase space closes rapidly as  the decay of $D^*\to D\pi$  becomes
kinematically forbidden at
$\xi=\xi_0\approx 1.03$. 
Consequently, the three-body cut in the on-shell partial-wave projected OPE potential turns into the lhc.  

An alternative view on the singularity structure of the problem can be gained by examining the trajectories
of the branch points in the complex energy plane of the  $T$-matrix,  see Eq.~(\ref{eq:lse}) and Fig.~\ref{ampl_cuts_E} for illustration. 
Starting at the physical pion mass with $\xi=1$ and $m_{\pi}<\Delta M$,
there are three branch points in the  amplitude near the $DD^*$ threshold (see left panel): the lowest in energy is the three-body cut 
starting from the $\pi DD$ threshold,  
followed by a pair of branch points in the second sheet of the complex plane in the propagator
 $G$,  related to the on-shell two-body $D^*D$ intermediate state. Here, the $\pi DD$ cut  generates the imaginary parts of the $D^*D$ branch points, related to each other 
 via the Schwarz reflection principle. As we increase the pion mass or, equivalently $\xi$, the
three branch points approach each other. At $\xi=\xi_0\approx 1.03$ the decay of $D^*\to D\pi$  becomes kinematically forbidden
and all three branch points coincide. If we  increase $\xi$ further, the number of branch points stays the
same, but their character changes (see right panel):  the left-hand cut from the OPE appears below the $D^*D$ threshold followed in
energy  by the  $D^*D$ branch point, which is now located on the real axis. Even higher up is the $DD\pi$ three-body cut. 
We emphasize that for all pion masses, besides the special case when $m_{\pi}=\Delta M$, the branch point of the three-body cut, as per Eq.~\eqref{Eq:branch_3body},
can only be reached if the incoming and the outgoing $D^*D$ state in the OPE potential is off shell (as discussed above, the three-body cut  can still occur in the on-shell potential for $m_\pi < \Delta M$ if the denominator in $D_1$ vanishes, but  only  at $k^2> k^2_{\rm rhc_3}$, and for $m_{\pi}=\Delta M$ the branch points of the on-shell and off-shell potentials coincide at $k^2=0$). 
While in the off-shell amplitude  this condition comes naturally from the off-shell potential
$V_{\alpha\beta}(E,p,p')$ itself, the
off-shell potential also enters the on-shell amplitude through iterations. 

Additionally, the lhcs from multi-pion exchanges are also present in
the amplitude, but are much more distant from the threshold and
therefore expected to have negligible impact on the process under consideration.

We now investigate the trajectory of the $T_{cc}^+$ pole as $\xi$ is varied, which is largely influenced by the
non-trivial motion of the branch points described in the previous paragraph.
The $T_{cc}^+$ pole at the physical pion mass can be interpreted
as a quasi-bound state -- a would be bound state of $DD^*$ if there were no three-body decay to $DD\pi$. 
When the pion mass increases, the $T_{cc}^+$ width    decreases
accordingly and the corresponding pole in the complex momentum plane
($k$-plane) approaches the imaginary axis, finally turning into a bound state. 
This is illustrated in  Fig. \ref{LO_pole_trajectory} -- see the zoomed plot in the left panel.  
Please note, however, that the imaginary part of the quasi-bound state energy is so small that   it is indistinguishable from zero  in the right panel of Fig.~\ref{LO_pole_trajectory}.
The proper bound state occurs  when the $T_{cc}^+$ pole, located below the $DD^*$ threshold,  coincides with the three-body threshold. This takes place at $\xi=\xi^\prime\approx 1.027$, which is just a bit smaller than $\xi_0$.
Therefore, there is a very narrow  range of $\xi_0>\xi> \xi^\prime$, where the three-body threshold is still below the two-body threshold, but the $T_{cc}^+$ is already stable. 
 By further departing from the physical point in terms of $\xi$, the bound state on the physical Riemann sheet  (RS-I) turns into a virtual state on RS-II.  
  The particular value of $\xi$ when this happens depends on the
  dynamics, namely on whether the LO or NLO potential is employed:  at
  LO of the chiral EFT expansion, the transition emerges at $\xi \approx 1.3$.
  A common feature of all settings is  the appearance of the second (lower-lying) virtual pole, the dynamics of which is interrelated with the location of the lhc from the OPE. The general pattern
   is illustrated in   Fig. \ref{LO_pole_trajectory} and is as
   follows: At some $\xi$, the second virtual pole occurs from under
   the lhc branch cut and goes along with the branch point until the
   first (upper) pole comes close. Then, both poles collide and the
   state becomes a resonance. The OPE plays a significant role, not
   only by  providing a repulsion, which would be absent in a pure contact formulation,  but also affecting the analytic properties of the 
   $DD^*$ scattering amplitude in a very nontrivial way. 
   The right panel of Fig. \ref{LO_pole_trajectory} shows
   the    behavior of the pole in the energy plane that 
   corresponds to the $k$-plane pole trajectory in the left panel.
   The cusps and the point where $E_{\rm pole} = 0$ indicate a change
   in the character of the pole, which transitions from a quasi-bound
   to bound, virtual and finally to a resonance state as $\xi$  increases. The point where $E_{\rm pole} = 0$ 
   corresponds to  the transition  from RS-I to RS-II. 
  Qualitatively, the pole trajectories described here are similar to those discussed, e.g., in Refs.~\cite{Hanhart:2008mx,Hanhart:2014ssa,Albaladejo:2012te,Pelaez:2010fj,Matuschek:2020gqe}, 
  though modified by the effects of  dynamical pions. These pionic effects induce the only hadronic contribution to the imaginary part of the $T_{cc}$ pole at the physical point and
  influence the $\xi$ dependence of the $T_{cc}$ pole as well as the second more distant virtual pole.

\subsection{\text{\rm NLO} results}\label{sec:pole_NLO}

\begin{figure}[t]
    \centering    
      \includegraphics[width=0.97\linewidth]{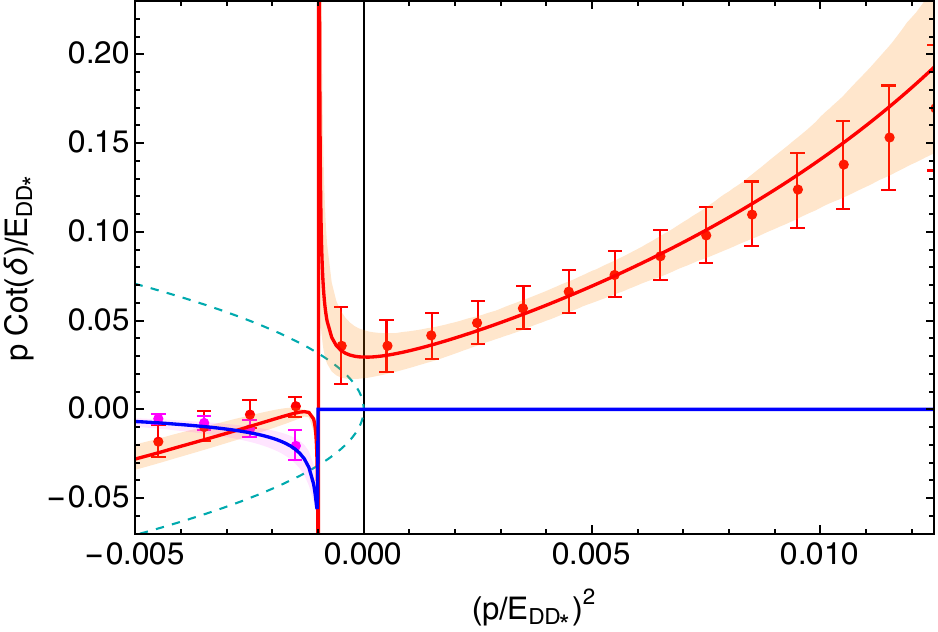}  
      \caption{ 
      Fit to the $DD^*$ scattering phase shifts, which were extracted previously in Ref.~\cite{Meng:2023bmz} from FV energy levels, including the lhc from the OPE.
          Red and pink dots  denote the real and imaginary parts of 
          $p \cot{\delta}$
          in the $^3S_1$ partial wave  from Ref.~\cite{Meng:2023bmz}. Red and blue
          lines are the results of the best fit  
          for the real and imaginary parts obtained in this work;  orange and  pink bands  represent the $1\sigma$ uncertainty. 
          Green dashed line corresponds to $ip = \pm|p|$ from unitarity. $E_{DD*} = M_D+M_{D*}$.
}
   \label{Fig:phase_shift_fit}
\end{figure}

 \begin{figure*}[t]
    \centering    
        \includegraphics[width=1\linewidth]{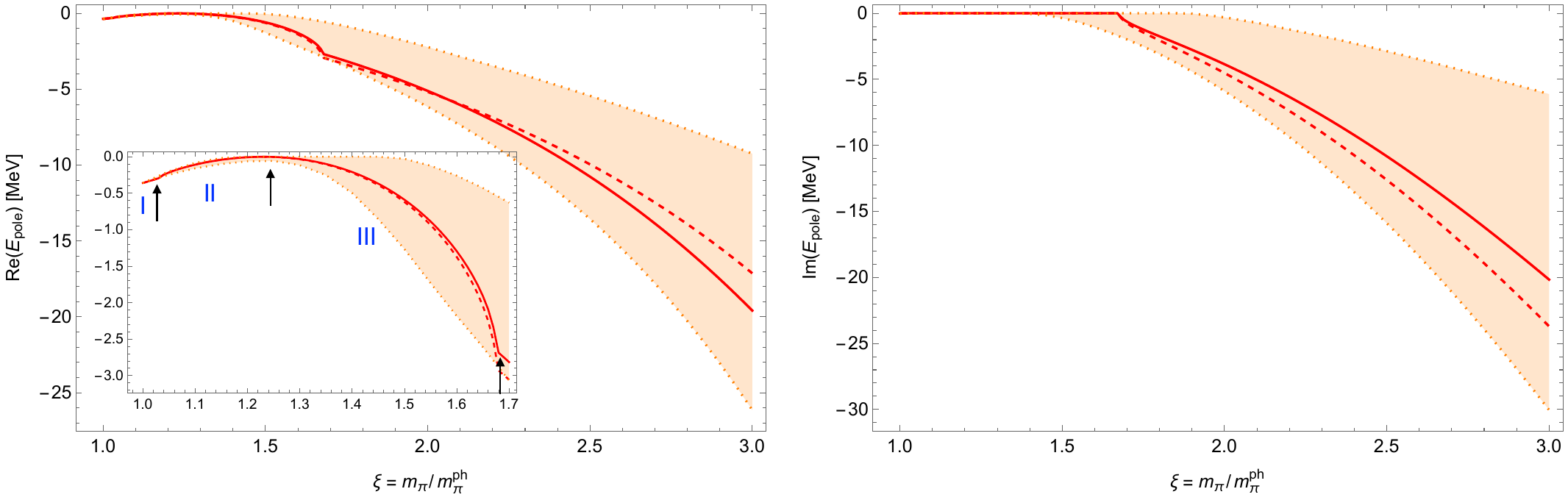} 
        \caption{
          The real  (left panel)  and imaginary (right panel) parts of the   $T_{cc}$ pole position at NLO in chiral EFT as a function of the pion mass. 
           The solid red line corresponds to the best fit at NLO with the $S$-wave OPE potential, while the orange band stands for 
          the $1\sigma$  error band estimated using  bootstrap.  The dashed red line corresponds to the best fit at NLO with the full OPE potential, including $D$ waves. 
          The inlay highlights the  behavior of the pole at lower pion masses, 
          where the transition from   quasi-bound (region I) to bound (region II), and then to virtual state occurs (region III), 
          as indicated by the arrows.  After  $\xi \approx 1.68$ the $T_{cc}$ becomes a resonance state and the Im part of the pole occurs in the right panel. 
}
   \label{NLO_pole_trajectory}
\end{figure*}

While the results at LO capture the  behavior of the pole trajectory qualitatively,  
in this section, we discuss how the results change when the higher-order contact interactions in Eq.~\eqref{full_NLO_contact_ansatz},
$C_2$ and $D_2$, are incorporated.  
First, we note that the experimental data~\cite{LHCb:2021vvq,LHCb:2021auc}
do not allow us to fix the momentum-dependent  ${\cal O}(Q^2)$ LEC
$C_2$ --- or stated otherwise, at the physical point there is a strong correlation between the values of $C_2$
and  $C_0$. Accordingly,
the EFT-based analysis of Ref.~\cite{Du:2023hlu} 
resulted in an excellent description of the experimental line shapes using $C_2=0$.  
On the other hand, this momentum-dependent short-range interaction was found to  
be important in Refs.~\cite{Du:2023hlu,Meng:2023bmz}  for understanding the results  
of Ref.~\cite{Padmanath:2022cvl} at $\mpi=280$ MeV  ($\xi\approx 2$).  
To extract the unknown parameters of the  short-range interactions at
${\cal O}(Q^2)$ --- specifically, $C_2$ and $D_2$  in
Eq.~\eqref{full_NLO_contact_ansatz} --- we   
use  the  phase shifts extracted in~\cite{Meng:2023bmz} from the lattice energy levels 
 at $\xi\approx 2$~\cite{Padmanath:2022cvl}.
 It is important to emphasize that the LECs extracted  in Ref.~\cite{Meng:2023bmz} cannot  be used directly in the current analysis, since  Ref.~\cite{Meng:2023bmz} approximates 
 the full $\pi DD$ Green function, as per Eq.~\eqref{Eq:Dprop}, with the static pion propagator.  While this approximation is    justified for analyzing lattice data at  $\mpi=280$ MeV, 
 the three-body cut emerging from the $\pi DD$  Green function is crucial for maintaining the correct analytic structure of the $DD^*$ scattering amplitude and 
 thus for  providing chiral extrapolations from the lattice data to the physical point. 
  Thus, we adjust the contact terms to the phase shifts with the main criterion that the resulting central curve and its uncertainty band should resemble those from Ref.~\cite{Meng:2023bmz}. 
 The results of our best fit, including  the $1\sigma$ uncertainty band, are shown in Fig.~\ref{Fig:phase_shift_fit}, along with the original phase shifts from Ref.~\cite{Meng:2023bmz}, and are in good agreement. 
 However, we note that we do not account for possible correlations between the input data points. This leads to a more conservative estimate of the propagated uncertainty.
 To propagate the uncertainties  from the original dataset into our
 calculations, we use the bootstrap procedure --- see, e.g.,  Ref.~\cite{Efron:1986hys} for details. 
 Specifically, we employ the orange band   in Fig.~\ref{Fig:phase_shift_fit} from Ref.~\cite{Meng:2023bmz} to randomly generate 1000 datasets,  assuming a Gaussian distribution.  Each of these simulated datasets is individually
 used to determine the best-fit parameters $C_2$ and $D_2$. 
 The resulting distribution of $\{C_2,D_2\}$   from these fits  is then propagated to estimate the  uncertainty of the results for the $DD^*$ scattering amplitude and the $T_{cc}$ pole position. 
  This uncertainty is associated with the statistical uncertainty of the finite-volume energy levels from Ref. \cite{Padmanath:2022cvl}. 
 
  \begin{figure*}[t]
    \centering     
       \includegraphics[width=0.7\linewidth]{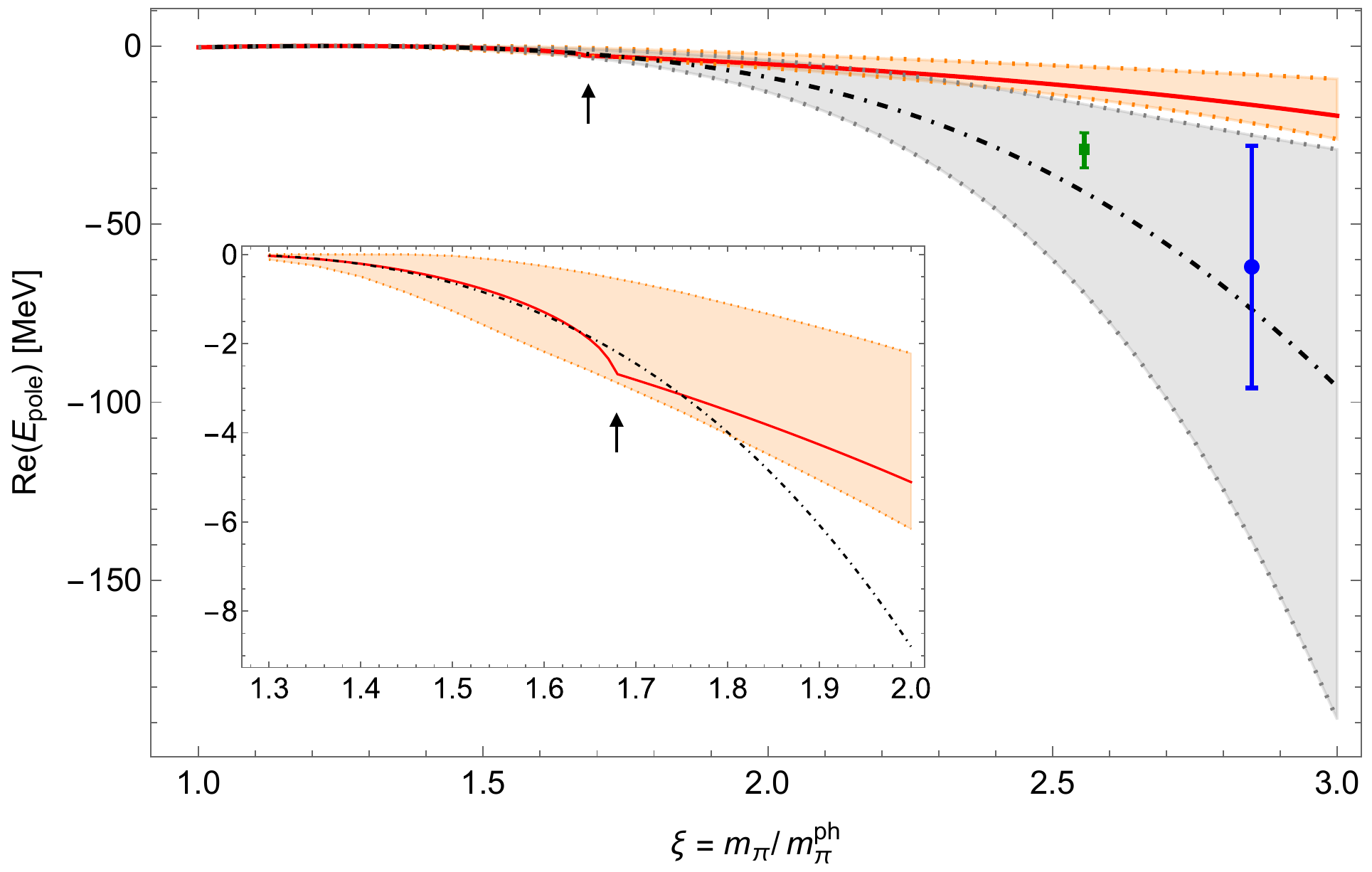}  
          \caption{
           Comparison    of the   $T_{cc}$ pole trajectory at NLO in chiral EFT   with that  in a pionless (contact)  theory (dot-dashed line).  
           Notation is the same as in Fig.~\ref{NLO_pole_trajectory}. 
          Only the real part is shown, as the pole in the contact theory is always real;  gray band indicates the statistical uncertainty of the contact EFT calculation propagated from fits to the lattice energy levels based on the ERE \cite{Padmanath:2022cvl}. The arrow indicates the pion mass ($\xi \approx 1.68$), after 
          which the pole in the pionful theory becomes a resonance state while  the pole in the pionless theory remains a virtual state. The blue data point corresponds to a virtual state extracted by the Hadron Spectrum Collaboration at $m_{\pi}=391$ MeV~\cite{Whyte:2024ihh},  while the green square corresponds to a virtual state  at $m_{\pi}\approx 348.5$ MeV, extracted  using the ERE parameters from Ref.~\cite{Chen:2022vpo}.
}
   \label{NLO_pole_trajectory_CT}
\end{figure*}

 In Fig.~\ref{NLO_pole_trajectory}, we show the results for the pole position at NLO in chiral EFT.  As a general pattern, one finds that the subleading short-range interaction provides additional repulsion, 
  causing both the transitions from bound to virtual state (when Re $E_{\rm pole}$ crosses $0$)
 and from virtual state to a resonance to occur at lower values of  $\xi$  compared to LO.  
 The $D$-wave contribution from the OPE is found to play a minor role in the pole trajectory, as seen by comparing the solid (pure $S$ wave) and dashed (with $D$ wave included) red lines in 
 Fig.~\ref{NLO_pole_trajectory}. 
Given the smallness of this effect, we neglect the $D$ wave components
when performing   uncertainty quantification. 
 This especially helps to simplify   the bootstrap procedure used to estimate statistical uncertainty of the results. 
  
In Fig.~\ref{NLO_pole_trajectory_CT}, we compare our NLO results with the pole trajectory obtained from a pure contact theory using only the contact potential from Eq.~\eqref{full_NLO_contact_ansatz} (see the dashed-dotted line). The LECs for the contact potential  were obtained by reproducing the physical value of the $T_{cc} $ pole position and the phase shifts of Ref.~\cite{Padmanath:2022cvl}, which were extracted from the finite-volume spectra using the Lüscher method and analyzed employing the ERE. We note that a recent lattice investigation of coupled $DD^*-D^*D^* $ scattering at $\mpi = 391 $ MeV (\(\xi \approx 2.85 $) \cite{Whyte:2024ihh}, also using the L\"uscher method
and ignoring the lhc, predicts a virtual state $62\pm 34$ MeV below the $DD^* $ threshold, fully consistent with the about 75 MeV virtual state predicted by our dot-dashed curve. 
In addition, the virtual state pole obtained  by extrapolating the ERE from Ref.~\cite{Chen:2022vpo} to the below-threshold region (see the green point) is also in line with the  prediction from our contact EFT.
On the other hand, it is evident that the additional repulsion from the longest-range OPE potential,  not included in Refs.~\cite{Whyte:2024ihh,Chen:2022vpo},  significantly impacts the results. 
Unlike the contact  trajectory, which  remains a virtual state with growing pion mass, our NLO pole position  transforms from a virtual state to a resonance at pion masses corresponding to $\xi \approx 1.7 $.

 The  behavior of the $T_{cc}$ pole trajectory with respect to the pion mass predicted in Figs.~\ref{LO_pole_trajectory} and \ref{NLO_pole_trajectory} is  consistent with 
a molecular nature of the $T_{cc}$ state.  Indeed,   the smooth transition of the pole trajectory from a bound state to a virtual state 
 as the light-quark mass changes is a distinguishing feature of a molecular structure~\cite{Matuschek:2020gqe}.

\section{Theoretical uncertainty and cross checks}
\label{sec:theor_uncert}

\subsection{Chiral truncation error}
\label{sec:truncation}

\begin{figure*}[t]
    \centering    
       \includegraphics[width=1\linewidth]{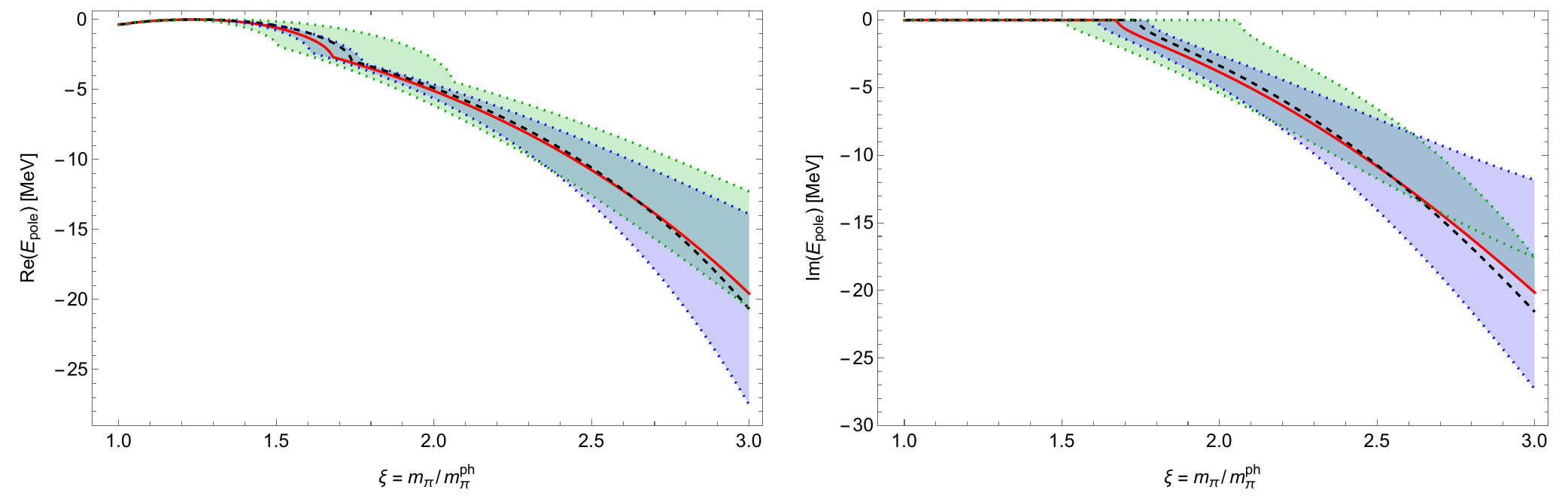}     
       \caption{ Theoretical uncertainty for the $T_{cc}$ pole position. The bands represent the truncation uncertainty of the chiral expansion for the $T_{cc} $ pole position at $\Lambda = 700 $ MeV when the ${\cal O}(Q^4) $ terms in Eq.~\eqref{Eq:V4} are included. The green band corresponds to the variation of the dimensionless constant $\alpha_4 $ in the range $[-1, 1] $, while the blue band corresponds to the variation of the dimensionless constant $\tilde{\alpha}_4 $ in the same range.
    Red  solid and black dashed lines  represent   the best fits for $\Lambda=700$ MeV  and  $\Lambda=500$ MeV, respectively. 
}
   \label{Fig:truncation}
\end{figure*}
In the previous section, we presented the results for the $T_{cc}$
pole trajectory and estimated errors by propagating the statistical uncertainty of the lattice data.  
In this section, we provide an estimate of the theoretical uncertainty which comes from 
the truncation of the chiral expansion and the cutoff dependence.  

The uncertainty associated with the truncation of the chiral expansion can be estimated by introducing the higher-order $\mathcal{O}(Q^4)$ terms, 
not explicitly included in the calculations so far, and evaluating their impact on the results.
The $\mathcal{O}(Q^4)$ terms used for the uncertainty quantification are  
\bea\label{Eq:V4}
V_{\text{cont}}^{(4)} = 
 D_4(\xi^2-1) (p^2 +{p'}^2) +\tilde D_4(\xi^4-1), 
\eea
while the effect of the $\mpi$-independent  $\mathcal{O}(p^4)$ contact term  is  
neglected, as this is consistent with  available experimental information at   $m_{\pi}=m_{\pi}^{\text{ph}}$    and lattice data at $m_{\pi} = 280$ MeV.
Here,   $ D_4$ and $\tilde D_4$ can be expressed as 
\bea\label{def_alpha_4}
  D_4= \frac{\alpha_4}{F_{\pi}^2}\left(\frac{m_{\pi}^{\text{ph}}}{\Lambda_\chi^2}\right)^2, \hspace{0.5 cm} 
    \tilde D_4= \frac{\tilde\alpha_4}{F_{\pi}^2}\left(\frac{m_{\pi}^{\text{ph}}}{\Lambda_\chi}\right)^4,  
\eea
 with $\Lambda_{\chi} \simeq 1$ GeV being  the chiral  symmetry breaking scale.  
Here,  $\alpha_4$  and $\tilde \alpha_4$ are   dimensionless prefactors  expected to be of the order of $1$ based  on naturalness, as discussed analogously in the $NN$ case~\cite{Epelbaum:2002gb}. 
Indeed, applying the same logic to the terms $C_2$ and $D_2$ in Eq.~\eqref{full_NLO_contact_ansatz}, we can rewrite them in the following form,
\bea\label{def_alpha_1}
C_0 = \frac{\alpha_0}{F_{\pi}^2}, \hspace{0.3 cm}   
   C_2 = \frac{\alpha_2}{F_{\pi}^2}\frac{1}{\Lambda_\chi^2}, \hspace{0.3 cm}   
   D_2 = \frac{\tilde\alpha_2}{F_{\pi}^2}\left(\frac{m_{\pi}^{\text{ph}}}{\Lambda_\chi}\right)^2,   
 \eea
  where the values of $\{\, \alpha_0, \,\alpha_2, \,\tilde\alpha_2\}$ from our best   fit to the data are  $\alpha_0\approx -0.12$, $\alpha_2\approx 0.20$ and $\tilde\alpha_2\approx 0.42$,  consistent with  expectations.

To estimate the impact of the $\mathcal{O}(Q^4)$ terms, we allow $\alpha_4$ and $\tilde{\alpha}_4$ to vary from $-1$ to $+1$. 
This range is considered conservative, given that the values of $\{\alpha_0, \alpha_2, \tilde{\alpha}_2\}$ are smaller. 
  The uncertainty associated with the truncation of the chiral expansion is shown in Fig.~\ref{Fig:truncation} (see the green and blue bands).  To  obtain these bands, we use the best fit parameters for $C_0, C_2$ and $D_2$ from our NLO results and supplement 
  the calculations with the higher-order potential  $V_{\text{cont}}^{(4)}$  from Eq.~\eqref{Eq:V4}. No refit of  the phase shifts at $\mpi =280$~MeV 
  was performed,  
   resulting in a more conservative estimate.  It is  reassuring that
   the spread in the results at $\mpi =280$~MeV is  quite natural, and
   it appears to be comparable in size to the statistical uncertainty 
  from Fig.~\ref{NLO_pole_trajectory}. 
    The resulting uncertainty grows with $\xi$, but it  remains
  comparable with the statistical error in Fig.~\ref{NLO_pole_trajectory}. 
  
In Fig.~\ref{Fig:truncation}, we also illustrate the cutoff dependence of the pole trajectory,  
by comparing   the results of the best fits for   two cutoffs: $\Lambda=700$~MeV and $\Lambda=500$~MeV.  
As expected, this dependence appears very mild after the refit, and falls well within the truncation error.

 Additionally, we verified that the change in $\text{Im} E_{\text{pole}}$ of the  $T_{cc}$  at the physical pion mass, due to the inclusion of the contact term $C_2$, with this LEC fixed  at  $m_{\pi}=280$~MeV, is very small after adjusting $C_0$ to reproduce $\text{Re} E_{\text{pole}} $, and remains well within the uncertainty estimated in Ref.~\cite{Du:2021zzh}.

 Finally, another potential source of uncertainty arises from the transition between the real world and the isospin limit at the physical pion mass. 
As discussed above, the parameter $C_0 $ was adjusted to reproduce the real part of the $T_{cc}$ pole position, $\text{Re} E_{\text{pole}} = -356 $ keV, 
as extracted from the coupled-channel analysis in Ref.~\cite{Du:2021zzh}. However, in the isospin limit, the $T_{cc} $ becomes slightly more bound. 
To estimate the impact of this effect on the pole trajectory, we adopt the following strategy:  (i) 
Starting from the contact coupled-channel framework of Ref.~\cite{Du:2021zzh} (see fit 1 in Tables I and II), we use the extracted value of $C_0$ to 
calculate the $T_{cc}$ pole in the isospin limit,
which gives    $\text{Re} E_{\text{pole}}\approx -940$~keV;
(ii) Relying on the fact that $\text{Re} E_{\text{pole}}$  at the physical pion mass is largely insensitive to  pion dynamics~\cite{Du:2021zzh}, we use this  binding energy as input in the full pionful framework in the isospin limit, recalculate $C_0$\footnote{Note that  we cannot directly use  $C_0$ from the coupled-channel {\it pionful} fit of~\cite{Du:2021zzh} in the isospin limit  for the following reasons: (i) unlike Du et al., we employ a formalism 
 with relativistic pions; (ii) in the full approach with pions, $C_0$ is known to exhibit a limit cycle, meaning it can rapidly change from  $-\infty$ to $+\infty$
as a function of the cutoff  \cite{Nogga:2005hy,Baru:2011rs}   
while the observables remain cutoff independent. 
Due to these reasons, taking $C_0$ from  one pionful framework and applying it in a different one could lead to large, uncontrolled systematic errors. 
This, however, is expected to work well in the  contact case.}, 
and find that the impact of this effect on the pole trajectory is very minor (just slightly larger than the effect of  cutoff variation shown in Fig.~\ref{Fig:truncation}). 
As expected for a more attractive potential, all transitions (quasi-bound to bound, to virtual, to resonance) occur at slightly larger pion masses.

\subsection{Comparison to other works}

Here,  we  briefly comment 
 on   other lattice calculations available in the literature and compare them  with our results. 
 In Ref.~\cite{Lyu:2023xro},  the HAL QCD method was utilized  to extract the
 $DD^*$ scattering potential at the pion mass $m_{\pi}=146.4$ MeV,  which was then employed to calculate the phase shifts above the 
 $DD^*$ threshold.  No visible signature of the OPE was observed. 
 Employing the ERE, the $T_{cc}$ pole was reported to be a virtual state with $k= (-8 \pm 8^{+3}_{-5}) i $~MeV corresponding to 
 $E_{\rm pole} = -59 \substack{^{+53\,+2} \\ _{-99\,-67}}$~keV, 
 where the errors represent  statistical and systematic uncertainties,
 in order.  Extrapolation of the HAL QCD potential to the physical pion mass, 
 neglecting three-body and isospin violating effects, led to a bound state but with a binding energy substantially 
 smaller than observed experimentally.
 
 The pion mass $m_{\pi}=146.4$~MeV corresponds to $\xi \approx 1.07$,  with the $DD\pi$ threshold  above the $DD^*$ threshold.  Thus the  
 three-body decay is already closed but the system experiences the left-hand cut from the OPE.   
 The pole  in our approach corresponds to a bound state with $k= (19 \pm 1) i$~MeV and $E_{\rm pole}= {-179\pm25 }$~keV. 
While our pole position is different to that of HAL QCD
quantitatively,   one may still conclude that the results are
qualitatively consistent, 
as a  small modification in the potential is typically sufficient to shift the pole from a bound to a virtual state.

 To summarize  this discussion, 
 the OPE has a significant effect on the imaginary part of the pole location of the $T_{cc}$ for the
 physical pion mass. 
 This width is primarily due to the hadronic decay of the   $T_{cc}$  to $DD\pi$, with the strength of the OPE controlled by the observable $D^* \to D\pi$ decay. 
 A small increase in the pion mass by about 7\% does not change the
 pion coupling significantly, so that the strength of the OPE at $m_{\pi}=146.4$~MeV 
 remains comparable to that at the physical point.
On the other hand, the effect of the OPE on the real part of the pole is not so dramatic for pion masses close to the physical point. Indeed, 
it follows from Fig.~\ref{NLO_pole_trajectory_CT} that for pion masses $\mpi \lesssim 230$~MeV ($\xi \lesssim 1.7$), 
very precise lattice simulations
would be required to quantitatively discriminate between the pure contact and pionful results. 
However, the impact of the OPE for $\mpi \gtrsim 230$~MeV is very significant since the repulsion generated by the OPE shifts 
the pole into the complex plane, creating substantial differences from contact results.

In Ref.~\cite{Collins:2024sfi},  the charm-quark mass dependence of the $T_{cc}$ pole position was investigated for a pion mass $m_\pi = 280$ MeV and several values of the heavy-quark mass.
While the phase shifts were extracted from the FV spectra using the L\"uscher method,  ignoring the left-hand cuts,  the $T_{cc}$ pole position was calculated from fits to the phase shifts in a framework 
involving the OPE.   The pion mass dependence of the $T_{cc}$ pole position was also briefly discussed but very qualitatively, without making quantitative predictions for the pole trajectory. 

 In Ref.~\cite{Chen:2022vpo}, elastic  $DD^*$ scattering in S-wave was investigated at $\mpi \approx 348.5$~MeV.  
However, the study was limited by using only a single lattice volume and having only one data point in the near-threshold region.  As a result of the ERE fits to the phase shifts obtained using the Lüscher method,  
the effective range parameters were presented
 -- the scattering length and the effective range. Using these parameters, we extracted the pole which turns out to be a virtual state about $-28^{+4}_{-5}$ MeV, as shown in Fig.~\ref{NLO_pole_trajectory_CT} by the green dot. 
While this result ignores the physics related with long-range interactions, it appears  consistent with our  prediction based on  contact EFT.

 Very recently,  a   lattice investigation of coupled $DD^*-D^*D^*$
 scattering at $\mpi = 391$~MeV  was presented in
 Ref.~\cite{Whyte:2024ihh}, finding a sizable coupled-channel effect.
 Using the Lüscher method and parameterizations of the coupled-channel
 amplitudes that ignore the OPE, two states were found in the energy plane: a virtual state about $62(34)$~MeV below the $DD^*$ 
 threshold and a resonance about ($49(35) + i 11(13)/2$)~MeV below the $D^*D^*$ threshold.  By employing a contact theory, we can indeed obtain a virtual state around $75$~MeV below the $DD^*$ threshold. However, this state transforms into a resonance with a substantially different pole position when the lhc effect from the OPE is included.
Further investigations are needed to investigate the impact of the lhc on the  coupled-channel transitions and the pole below the $D^*D^*$ threshold.

\section{Summary and conclusions}
A chiral EFT approach is utilized to establish a continuous connection between the $T_{cc}$ pole position at unphysical pion masses from lattice simulations and the physical world.
The effective potential, extended up to next-to-leading order (${\cal
  O}(Q^2)$), includes  three contact terms and the longest range
one-pion exchange potential that incorporates
all relevant scales associated with the three-body $DD\pi$ cut and the
left-hand cut.
A proper treatment of these scales is necessary to fulfil analyticity and unitarity of the
$DD^*$ scattering amplitude
when the pion mass deviates from its physical value.
The contact terms are adjusted  to reproduce the $T_{cc}$ pole position at the physical pion mass  and lattice finite volume spectra at $m_{\pi} = 280$~MeV.  
This allows us to predict the trajectory of the $T_{cc}$ pole as a function of
the pion mass up to $m_\pi \sim 3 m_\pi^{\rm ph}$.  
To propagate the statistical uncertainty from the lattice data at $m_{\pi} = 280$ MeV to the pole position at different pion masses, 
the bootstrap method is employed. Additionally, the truncation uncertainty of chiral EFT is estimated by including two higher-order (${\cal O}(Q^4)$) contact terms based on naturalness.

The resulting position of the pole transitions from a   quasi-bound to  a bound, a virtual and a resonance state as the pion mass increases, indicating the molecular nature of the $T_{cc}$.  
Observing the effect of the OPE on the pole for pion masses slightly above the physical value is challenging and may require very precise lattice simulations.  However, for $\mpi \gtrsim 230$~MeV, 
the presence of the OPE is evident as the repulsion it generates shifts the pole into the complex plane, leading to  substantially different results from those obtained without pions.

\bigskip

\bigskip

This work was supported in part by the MKW NRW under the funding code NW21-024-A, by DFG and NSFC
through funds provided to the Sino-German CRC 110 “Symmetries and the Emergence of Structure in QCD” (NSFC
Grant No. 11621131001, DFG Project-ID 196253076 - TRR 110), by ERC
NuclearTheory (grant No. 885150), by BMBF (Contract No. 05P21PCFP1), and by
the EU Horizon 2020 research and innovation programme (STRONG-2020, grant agreement No. 824093).

\bibliographystyle{elsarticle-num} 
\bibliography{Tcc_refs} 

\end{document}